# Data Work in Memory Institutions: Why and How Information Professionals Use Wikidata

RIYA SINHA, School of Information, The University of Texas at Austin, USA
AMELIA ACKER, School of Communication & Information, Rutgers, The State University of New Jersey, USA
HANLIN LI, School of Information, The University of Texas at Austin, USA

Wikidata, an open structured database and a sibling project to Wikipedia, has recently become an important platform for information professionals to share structured metadata from their memory institutions–organizations that maintain public knowledge and cultural heritage materials. While studies have investigated why and how peer producers contribute to Wikidata, the institutional motivations and practices of these organizations are less understood. Given Wikidata's potential role in linking and supporting knowledge infrastructures and open data systems, we examined why and how information professionals in memory institutions use Wikidata as part of their organizational workflow. Through interviews with 15 participants, we identified the three archetypal roles of Wikidata users within memory institutions–providers, acquirers, and mutualists–and the different types of contributions that these institutions bring to Wikidata. We then explored potential collaboration opportunities between memory institutions and other volunteers in Wikidata, discussed the value of the data work conducted by these professionals, and examined how and why they track their contributions. Our work contributes to the wider discussions around collaboration and data work in CSCW by 1) studying the motivations and practices of information professionals, their differences from those doing volunteer work, and opportunities for the Wikidata community to promote more collaborative efforts within memory institutions and with other volunteers and 2) drawing attention to the important data work done by memory institutions on Wikidata and pointing out opportunities to support the contributions of information professionals.

## 1 Introduction

Wikidata, an open structured database and a sibling project of Wikipedia, has become a prominent platform for memory institutions (such as libraries, archives, and museums) to share structured metadata about their holdings [8, 52]. Wikidata offers memory institutions the opportunity to share and update information about their collections [27, 57], supporting the "collections as data" movement that advocates for responsible computational use of collections [44]. For example, Candela

---

Authors' Contact Information: Riya Sinha, riyasinha@utexas.edu, School of Information, The University of Texas at Austin, Austin, TX, USA; Amelia Acker, amelia.acker@rutgers.com, School of Communication & Information, Rutgers, The State University of New Jersey, New Brunswick, NJ, USA; Hanlin Li, lihanlin@utexas.edu, School of Information, The University of Texas at Austin, Austin, TX, USA.

et al. noted that memory institutions can use Wikidata to enrich their collections by linking and extracting data objects such as standardized names or geographic locations [19]. The WikiProject *PCC (Program for Collaborative Cataloging) Wikidata Pilot* has 79 member institutions whose employees "create trusted metadata and support its use and reuse by global communities" on Wikidata [8]. As Wikidata is a machine-readable resource available to all, the contributions of information professionals from these memory institutions have the potential to address critical knowledge gaps in downstream technologies, including Wikipedia, search engines, large language models, and many other "invisible machines" that reuse Wikidata data [71]. Strengthening Wikidata's integration with memory institutions, as both users and contributors, has the potential to enrich existing knowledge infrastructures and open data systems.

We currently lack an in-depth understanding of why and how information professionals from memory institutions work with Wikidata and its role in their organizations. Although studies on libraries and education have examined the adoption of Wikidata by memory institutions, they tend to focus on the utility of Wikidata rather than the motivation, practices, and processes of information professionals [24, 52]. Social computing researchers have studied individual volunteers' motivations and practices on Wikidata [41, 71]; however, information professionals are likely to have additional or completely different motivations due to their training, professional background, and organizational goals. For example, the practices of information professionals may differ from those of individual contributors driven by personal interest [71]. The former's responsibilities for creating and managing accurate and high-quality representation information for subject-specific collections in their institutions [26] may also be shaped by their professional ethics that promote access to information and espouse digital openness to cultural heritage collections [1, 2, 4].

The differences between how information professionals and individual volunteers use Wikidata are unclear [23]. If left unaddressed, these unknowns prevent memory institutions, Wikidata communities, Wiki developers, and researchers from supporting information professionals' valuable contributions to Wikidata and the broader knowledge infrastructures. Memory institutions host massive caches of information, data, and metadata from their unique and historical collections in the public interest[52]; These collections have the potential to substantially enrich Wikidata and downstream technologies such as Siri, Alexa, and large-language models[21]. For the Wikidata community, "harnessing the wisdom of crowds"[33], especially the inputs and expertise from memory institutions, is invaluable for addressing knowledge gaps, ontology engineering [41], and preserving cultural heritage [76], among other possible benefits.

To explore the motivations and use of Wikidata among information professionals, we conducted interviews with 15 professionals from 12 distinct libraries, archives, and museums. Our interview protocol involved questions about why and how information professionals currently contribute to and use Wikidata as part of their occupational responsibilities in memory institutions. We focused on their approaches and data work practices to inform opportunities to support better integration of memory institutions with Wikidata.

Our research identified three user archetypes (Figure 1) of Wikidata in memory institutions: 1) providers: organizations that primarily upload their data to Wikidata, mainly driven by goals of sharing and broadening access to their collections, 2) acquirers: organizations that pull in data from Wikidata to enrich their local collections (e.g., adding birth dates to authors, or to introduce Wikidata links as external references for their visitors), and 3) mutualists: organizations that upload data to and download data from Wikidata for sharing and managing their records. We found that information professionals contribute data on various subjects, for example, data about books, authors, file formats, and software. We also identified distinct collaboration processes ranging from expertise-driven group efforts to collective experimental exploration. Although Wikidata

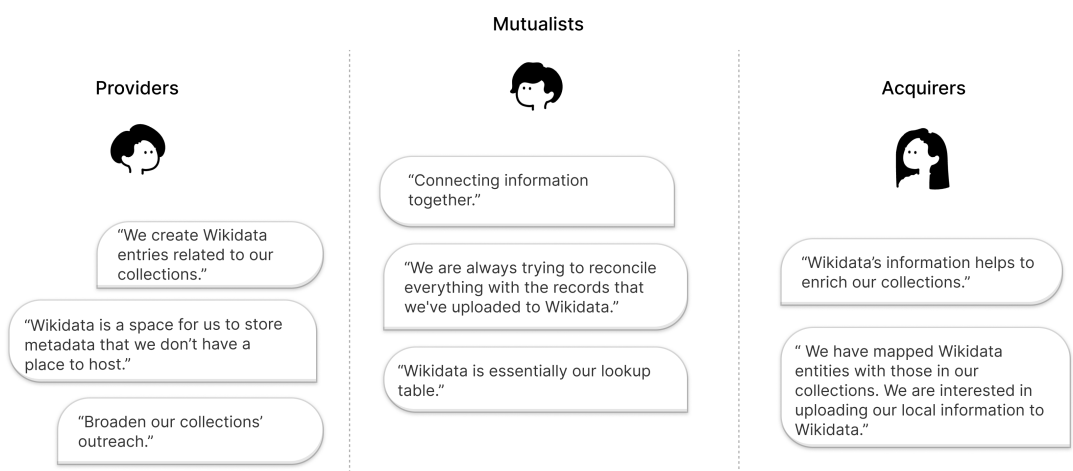

Fig. 1. The Archetypes: The above figure represents the three archetypes —providers (those who upload new data to Wikidata), mutualists(those who both upload new data and download data from Wikidata), and acquirers (those who only download data from Wikidata)

supports collaborative peer production, we saw limited instances where information professionals collaborated across institutional boundaries. Further, while many information professionals regularly track their collective contributions to Wikidata, the impact of these contributions on the broader knowledge infrastructures consisting of Wiki systems, search engines, and digital collection systems remains unclear.

The contribution of our work is two-fold. First, our findings about the archetypes of Wikidata integration in memory institutions point to opportunities for providers and acquirers to take full advantage of Wikidata by becoming mutualists. Accordingly, we make design, policy, and education recommendations based on these archetypes. Second, we study the motivations and practices of information professionals in depth and argue for Wikidata's potential for supporting collaborative data work among memory institutions and helping them uphold, steward, and maintain knowledge infrastructures and open data systems.

## 2 Background
### 2.1 Collections as Data in Memory Institutions

The origins of datafication in memory collections of cultural heritage can be located in the long history of metadata development and computerized collections management through metadata standards like MARC (MAchine-Readable Cataloging). These standards have been used since the 1960s in libraries and data archives to exchange metadata records about collections [14, 26]. Today, collections from memory institutions, even museum objects and analog archives in special collections, are considered data. This shift is partly due to what Kitchin calls the data revolution of our contemporary age, where all forms of human activity are perceived as both data and resources [32]. More recently, the conception of institutional collections as data has been spurred by the multi-year Mellon-funded "Collections as Data Project" led by Thomas G. Padilla [45].

Collections are increasingly digitized, and full-text access is offered online. Linked open data infrastructure is a set of design principles for sharing machine-readable interlinked data on the web with free and open usage (of which Wikidata is an example) that enables access and reuse of data across institutions. Memory institutions view linking open data and publishing data online as essential for promoting discoverability. For example, linking MARC records to objects from digital collections or publishing bibliographic records to the Digital Public Library of America enhances access to these collections from other sources. Moreover, linked open data initiatives such as the California Digital Library, Europeana, or HathiTrust foster further reuse of digital collections because collections' data (and metadata) remain reusable across contexts. Candela et al. showed that, while memory institutions have traditionally provided access to resources through siloed catalogs, with linked open data, they can improve interoperability and facilitate access via computational means [20]. The reuse of digital data is a critical component of the collections-as-data approach, where making data machine-actionable allows collections' metadata to support a range of research activities, from text mining and mapping to data visualization [20].

The "collections as data" framework provides an important lens for understanding the integration of Wikidata with memory institutions. As memory institutions increase their access to collections via open data technologies, their contributions are reconfigured, not just as machine-readable data, but transformed into new online knowledge networks. Candela et al. and others note that memory institutions can use Wikidata to enrich their own collections by linking and extracting data objects such as standardized names or geographic locations [19]. The present study examines whether and how information professionals leverage Wikidata as an external source of information and a platform when managing their collections.

## 2.2 Information Professionals as Information Maintainers and Their Roles in Open Data

Information maintainers, including archivists, librarians, and data managers, are information professionals who ensure that digital infrastructures remain functional. This labor is often undervalued [55]. The turn toward conceiving collections as data has reshaped how memory institutions approach access to their archives and collections by emphasizing the potential for computational access and integration of digital infrastructure, potentially increasing online access [45, 46]. By treating collections of cultural heritage as data, institutions can transform their cataloging and collection metadata into more machine-actionable data for further reuse [9, 36]. However, this growing acceptability of using cultural heritage collections in computational contexts, particularly for AI, brings challenges. This is especially true for those information professionals responsible for creating and managing the representation information and metadata for their collections, such as archivists and metadata librarians[21]. As a professional archivist and researcher, Tansey highlights that the role of such interventions providing open access online is often overlooked, leading to the invisibility of the archivists' labor [62]. Archivists, librarians, and data workers are frequently marginalized within the institutional hierarchies that manage collections because of their work as "invisible technicians" [38, 58]. This present study is a response to calls from information professional communities for further examination and recognition of the significance of the maintenance work for which information professionals are responsible [35].

## 2.3 Wikidata

Wikidata has a knowledge base of hundreds of millions of entities open to anyone to access and edit. Unlike Wikipedia, which is highly descriptive and detailed, Wikidata uses a structured format to represent knowledge. Entities in Wikidata have associated properties, and each property is paired with at least one label. For example, the "Location" property of the entity for Mona Lisa has the label "Louvre Museum", where the painting is stored. Each entity on Wikidata contains an item page describing its properties and a discussion page that contributors can use to discuss changes, removals, and reversals.

Another key distinction between Wikipedia and Wikidata is that Wikidata is language independent. Wikipedia consists of different language communities, each creating its own pages and page content. As a result, the same concept can be described differently across different language versions [12]. In contrast, each Wikidata entity has only one version, but its properties (e.g. "Location") and labels may be translated into different languages by editors. Labels in different languages may be shown all together on the entity's page to describe the entity. A 2017 study shows that English is the most common language in both labels (11%) and properties (4%), surpassing all other languages [30].

Compared to Wikipedia, Wikidata is a less-studied subject, likely because it is a newer project (started in 2012). Existing research studies have focused on assessing the quality of Wikidata and identifying potential areas of improvement. Shenoy et al. highlighted that revision histories and constraint verifications can function as indicators of the quality of Wikidata entities [61]. Focusing on references, Amaral et al. found that most Wikidata references are easy to access, relevant, and authoritative [11]. Zhang and Terveen investigated the gender representation of Wikidata entities and found that only 22% of the human entities are about women [72]. In a recent mixed-methods study, Weathington and Brubaker examined Wikidata properties that encode queer identities, such as sexual orientation, sex, or gender, highlighting the social and technical frictions between people of diverse backgrounds, views, and understanding of queer terminologies and identities when documenting on Wikidata [68].

Editing on Wikidata is done in various forms. Users can manually edit the properties of any entity and provide appropriate references. It is also possible to use automation tools such as OpenRefine to batch upload contributions. Wikidata also supports the cautious use of editing bots [5], and human users can create bot accounts, seek community approvals, and use them to automate edits, provided they comply with Wikidata community guidelines. Researchers have studied and developed tools to promote batch uploading of information to Wikidata. In a 2019 study, Kaffee et al. found that bots together have the highest edit count on Wikidata [29]. Tanon et al. programmed the Primary Sources Tool and released it as open-source software to facilitate the transfer of the Freebase dataset onto Wikidata [48].

Wikidata is a valuable information and data source for Wikipedia and other Wikimedia Foundation projects and it further supports search engines and other downstream technologies such as large language models and knowledge graphs [22, 37, 39, 59, 67]. Additionally, Alrashed et al. developed Wikxhibit, a tool that allows a user to author plain HTML to display any Wikidata entity easily with cross-references to other materials from the web [10].

## 3 Related Work
### 3.1 Data Work and Sharing

Data work refers to "any activity related to creating, collecting, managing, curating, analyzing, interpreting, and communicating data"[16] and is a core area of interest in the CSCW community. Researchers have emphasized different human activities involved in the data life cycle to support knowledge and memory construction[16, 49]. Data sharing is another subject of study for researchers interested in understanding humans' role in transferring data. Information professionals' practices around Wikidata can be viewed as data work through data creation, as they produce new data for either Wikidata or their local institutional databases. These practices can also be viewed as data sharing because they are transporting data from Wikidata to local databases or vice versa. Below, we discuss relevant studies from both areas and their connections with our work.

Researchers have studied how data creation and documentation happen in different use cases to support collaboration. For example, in the healthcare space, a multitude of data work has been carried out, such as improving communication between patients and doctors by infrastructuring telehealth medicine, understanding the perspectives of front-line health workers and their needs in data collection, and valuing data across stages of data processing [13, 28, 40, 63]. In a qualitative study, Bossen and Pine describe the successful collaboration between Clinical Documentation Integrity Specialists (CDIS) and an Artificial Intelligence (AI)-embedded software. The resultant data work made medical documentation more accurate, consistent, and complete [15]. These studies highlight the importance of data work in collaborative environments. Our study extends this line of literature by contributing to an understanding of the data work practices of information professionals who contribute to Wikidata.

Researchers have also studied how data sharing occurs. Vertesi and Dourish argued that the value of data is not fixed but shifts depending on the context of its production and the particular infrastructures that support its collection, cleaning, and dissemination [65]. They highlight that the meaning and value attributed to data are often shaped by the specific practices and norms of the communities involved in producing it. This perspective suggests that data sharing among information professionals is not simply about making data available but involves carefully considering the labor and resources involved in creating and maintaining data for reuse. Similarly, Neang et al. examined how scientists manage and integrate data across varied sources [42] and highlighted the significant articulation work and coordination required for data sharing. Additionally, within the Wiki ecosystem, Yu et al. studied the stitching work involved in sharing data between Wikipedia and Wiki Commons and identified key challenges due to the inherent differences in how the two platforms are structured [70]. Our study extends the literature by contributing a closer examination of open data sharing practices by information professionals [35, 51].

### 3.2 Motivations and Roles in Peer Production

Scholars have extensively studied contributor roles and participation in peer production projects. From open-source software projects [34], to citizen science initiatives [54], to Wikipedia [47], scholars have studied how personal motivations of active contributors shape their subsequent behaviors on peer production platforms. For example, Trinkenreich et al. focused on Open Source Software contributors and found that community-centric and project-centric members engage in different activities [64].

Within the Wiki space, Wikipedia has been a well-studied subject. Many Wikipedia editors are motivated by a mixture of intrinsic factors, such as altruism, and extrinsic factors, such as a sense of community [25, 43]. Through qualitative and quantitative means, researchers have also constructed distinct social roles to categorize Wikipedia editors, such as power editors, copy editors, peripheral editors, and novices [18, 47, 69, 75].

In contrast, studies investigating the motivations and roles of Wikidata contributors are relatively scarce, likely because Wikidata was launched more recently. Zhang et al. found that although many Wikidata contributors are motivated by altruism like Wikipedia editors [25, 43], some also have unique motivations, such as interest in creating data structures and organizing data [71]. Researchers have also examined the roles of Wikidata contributors based on their interests and activities. Müller-Birn et al. categorized Wikidata contributors into six groups: reference editors, item creators, item editors, item experts, property editors, and property engineers [41], while Piscopo et al. focused specifically on users' contribution to the ontology of Wikidata and identified two roles: contributors and leaders [50].

Our work extends the line of inquiry on the motivations and roles in peer production, centering on information professionals, given their potential to enrich the global knowledge base. More specifically, we examine how information professionals fulfill their professional commitments to the preservation of cultural heritage materials, public access, and an ethics of care by contributing to Wikidata [1–4, 17]. We also expand our focus by including information professionals who do not make direct contributions but download data from Wikidata [23, 24]. In doing so, our work contributes to a comprehensive understanding of Wikidata's role in memory institutions and the corresponding motivations and roles of information professionals.

A unique characteristic of the contributions of information professionals to Wikidata is that they can be seen as paid or sponsored contributions [1]. Wikidata is a space where paid work by information professionals and unpaid work by volunteers coexist. In this aspect, Wikidata is similar to OpenStreetMap, which accepts corporate and volunteer contributions. A recent study has shown that contributions made by corporate map users such as Apple Maps have become a substantial force on OpenStreetMap [66]. Open-source software projects are another space where unpaid and paid work occur together [53, 73, 74]. Such prior work has raised questions about the potential implications of paid contributors on the sustainability and focus of these peer production projects and whether paid contributors have an outsized impact on the community's focus and production. Our work engages with this line of research; we highlight how some information professionals' goals and practices differ from those of volunteer contributors, and how to shape the two groups' contributions to be complementary to each other.

## 4 Methods

With a snowball sampling approach, we recruited participants from LD4[2] (a community of information professionals interested in linked data), the iCaucus network of Information Schools, and the National Digital Stewardship Alliance. We selected participants based on two criteria: 1) participants are currently employed at a memory institution such as an archive, a library, or a museum, and 2) participants have reported experiences with contributing to Wikidata or using Wikidata as part of their duties in their current role. Table 1 provides an overview of the background and title of each participant, represented through a unique participant identifier that begins with the letter "P" followed by a number. Our institution's IRB office approved this study.

We recruited 15 participants from 12 unique memory institutions. Three participants contributed to Wikidata as part of the Wikidata PCC pilot project, and three participants from Africa and Oceania started their Wikidata journey after being exposed to the Wikimedia Foundation's outreach efforts around Wikipedia (e.g., Wikimedian in Residence and educational workshops). We conducted semi-structured online interviews with participants. The interview protocol was designed to elicit information about participants' practices when working with Wikidata (see the appendix for the interview protocol). We also examined the specific tools and workflows they utilize when contributing to Wikidata. For example, participants were asked to explain their use of editing tools, metadata, and collaboration techniques with colleagues in their organizations. We also asked participants to reflect on how they assess the impact of their work on Wikidata. Interviews ran from 42 to 70 minutes and lasted approximately 56 minutes on average. The participants were compensated with a $30 virtual gift card. The interviews were transcribed using an automated transcription service. After receiving the automated transcripts, the research team reviewed and manually corrected the transcripts to ensure accuracy.

Once the transcripts were prepared, we employed templated qualitative coding to analyze the data [56]. First, the research team developed a codebook through an iterative process. The initial list of codes was derived from our research inquiry and the interview protocol, focusing on categories such as "data curation practices," "collaboration," and "institutional integration and impact tracking." The three coders then applied the initial codebook to five interview transcripts. During this first round of coding using AtlasTI software, we refined the code themes and definitions. For example, the code for "data curation practices" refers to instances of how and why information professionals use Wikidata and comments about collaboration within the institutions. P3 mentions that they *"create authority file [and] authority names for these place names so that we can kind of tag all of the collections in the archive with place names.",* and this was coded as a data curation practice. The code "Challenges and Barriers" refers to mentions of challenges to using Wikidata in memory institutions and barriers to contributing to Wikidata. Many participant quotes that referred to a "steep learning curve for Wikidata" were coded as challenges and barriers. We confirmed intercoder reliability by comparing coded sections and discussing discrepancies in interpreting the same set of two interviews. We made revisions to the codebook based on our discussions and arrived at a finalized list of ten codes (Table 2 in the Appendix).

---

[1]https://www.wikidata.org/wiki/Wikidata:Disclosure_of_paid_editing
[2]https://sites.google.com/stanford.edu/ld4-community-site/home

Table 1. The table details an overview of participants' background, showcasing participant ID, archetype (refer to 5.1), title, degree, year the degree was granted, and organization. *, **, and *** denote participants from identical organizations.

| ID | Archetype | Title | Degree | Year granted | Organization |
| --- | --- | --- | --- | --- | --- |
| P1 | Provider | Head of Technical Services | MLIS | 2012 | Library |
| P2 | Provider | Head of Metadata and Cataloging | MLIS | 2014 | Library |
| P3 | Provider | Cataloging Specialist | MLS | 2013 | Library |
| P4 | Mutualist | Digital Outreach Manager | Masters | 2010 | Museum |
| P5* | Mutualist | Research Specialist | PhD | 2023 | Library |
| P6 | Mutualist | Head of Technical Services | MS | 2010 | Museum |
| P7 | Provider | Metadata Specialist | MLIS | 2018 | Library |
| P8** | Provider | Head of Digital Preservation | MLIS | 2016 | Archive |
| P9* | Mutualist | Head Librarian | PhD | 2018 | Library |
| P10 | Provider | Digital Assets Librarian | MA | 2020 | Library |
| P11 | Provider | Librarian | MLIS | 2009 | Library |
| P12 | Acquirer | Head of Technical Services | BS Software Engineering | | Museum |
| P13*** | Acquirer | Archivist | MLIS | 2015 | Archive |
| P14*** | Acquirer | Archivist | MLIS | 2017 | Archive |
| P15** | Provider | Archivist | MLIS | 2017 | Archive |

Following the first coding round, the team applied the finalized codebook to the interview corpus. Each team member wrote reflective memos after coding each transcript, noting any significant or unique patterns in the participants' contributions to Wikidata, institutional norms, or other notable observations to discuss with the team. Post-coding, memos were used during our team meetings to guide further analysis and thematic development. The coding process was completed once all transcripts had been fully coded and all significant themes had been identified and discussed.

During the final analysis stage, the team reviewed the coded segments across the entire interview corpus to identify key themes, connections, and generalizable observations. Team members synthesized these findings into comprehensive code summaries across the entire interview corpus. We identified and discussed emergent patterns we saw in participants' practices and then constructed three user archetypes– providers, mutualists, and acquirers. These archetypes are based on how information professionals currently use and contribute to Wikidata. We also conducted member checks with participants to confirm these archetypes.

## 5 Results

Below, we first provide the three user archetypes of information professionals based on the approaches they take to adopt Wikidata in their memory institutions–providers, acquirers, and mutualists. We then focus on providers and mutualists, those who contribute to Wikidata, and describe their practices and challenges, before unpacking how they leverage Wikidata as a tracking apparatus.

### 5.1 Providers, Acquirers, and Mutualists, and their Motivations

We identified three user archetypes of Wikidata representing distinct institutional approaches toward Wikidata: providers, acquirers, and mutualists.

**Providers** are information professionals from institutions that focus on uploading their data and metadata to Wikidata. The interaction between providers and Wikidata is one-way. Providers may query the data they have contributed in the aggregate to review their past work or gain an understanding of their aggregate contributions; however, they do not download data from Wikidata to their local collection systems. Our sample of providers included university libraries, a city public library serving a large urban population, and a national government archive, as mentioned in 5.1.1; all of these provider organizations have a mission to share their collections more broadly.

**Acquirers** can be seen as the opposite of providers because they focus on downloading Wikidata data rather than uploading new data. The information acquired from Wikidata is used to enrich memory institutions' collections. Acquirers first create a mapping of entities in their collection with those available on Wikidata and then download information about these entities from Wikidata to their collections. Acquirers may contribute their institutions' identifiers to Wikidata as properties (e.g., an identifier for a corporate name used by the National Library). Additionally, both acquirer organizations in our participant pool included Wikidata links in their access systems for visitors to explore.

While acquirers do not currently contribute data other than identifiers to Wikidata, this mapping paves the way for them to upload information about entities in their local collections to Wikidata in the future. Indeed, P12, P13, and P14 saw great potential in leveraging their existing mapping of entities to upload information that is not present in Wikidata. P13 and P14 made plans to create new pages for the entities in their collections that do not have a matched Wikidata page. In P14's words, "*If we could highlight people who are underrepresented in our collections, that would be an amazing way to contribute back [to Wikidata]*". However, these acquirer organizations faced resource and expertise constraints and did not further their integration with Wikidata.

**Mutualists** are information professionals from institutions that both upload new data to and download data from Wikidata. Mutualists in our sample started as providers by contributing their collections to Wikidata. Through this process, they produced a mapping of the entities in their collections and Wikidata entities. As with acquirers, mapping data resources enables mutualists to download relevant information about the collections' entities from Wikidata to add to their local databases.

Generally, acquirers and mutualists trust the information they extract from Wikidata and confidently reuse it in their collections. However, some may manually review the information extracted from Wikidata to ensure the descriptions comply with their professional standards or in-house cataloging policies. In this process, P13 and P14 identified a controversial figure in their collections whose descendants were major benefactors of their organization. These information professionals had to deliberate on balancing the accuracy of representation with the reputation management of this subject documented in their archive. For mutualists and providers, contributing to Wikidata can be a significant part of their job responsibilities. P1 shared that 90% of their job is to contribute to the work of researchers and students at their university to Wikidata to promote their scholarship. P2 and P4 also stated that their Wikidata work is a part of their job description. For example, P2 said *"My occupation is cataloger and Wikidata editor. So I do this as part of my work. I don't do it in my spare time when I'm not working."*

Below, we describe the shared and distinct motivations of providers, acquirers, and mutualists identified in our interviews with these information professionals.

*5.1.1 Open Knowledge and Access.* Sharing knowledge is a key motivation for providers and mutualists to integrate Wikidata with their institutions. Wikidata allows providers and mutualists to share their records and metadata with the public, creating a new venue for audience outreach. For some organizations, sharing knowledge is part of their organizational values, and Wikidata, given its open nature, becomes a natural tool to adopt. P2 from a public university in the United States, explains that *"Libraries are really into open scholarship, open software tools, open things. So Wikidata appealed to us."*

Some provider participants also compared Wikidata with other linked data infrastructures and discussed the openness of Wikidata as a unique characteristic. Unlike other proprietary infrastructures, any data they upload to Wikidata will stay available to everyone who wishes to reuse it. P7 in particular expressed the importance of openness in her work:

> *There are many linked data initiatives and projects, and something that is appealing about Wikidata is [that] it's kind of sustainable and maintained. It's a community infrastructure that won't disappear and is not owned by private companies... Everything we put on Wikidata, we can reuse; our peer institutions can reuse.*

Wikidata's integration with other open Wiki systems, such as Wikipedia and Wiki Commons, further strengthens information professionals' motivation to use Wikidata to share knowledge. P4 came across Wikidata while contributing images to Wiki Commons and creating structured data about specimens, collectors, and locations attached to those images. She found Wikidata helpful in supporting the data her team shares on Wikipedia and Wiki Commons.

Improving access is a direct result of sharing knowledge and is also a key motivation for providers and mutualists. They see their contributions to Wikidata as a patch to missing links or holes in existing access infrastructures. By creating links between Wikidata and items in their collections, providers make it easy for audiences to search for, locate, and access records. P7, from a city public library, described her organizational mission for access as a key motivating factor for adopting Wikidata: *"Because I'm working in a public library, we have such a public orientation and so many of our strategic directions are centered around access. Making sure that if we have materials, people can find them is a lot of the work that we do"*. In another case, P3 stressed how Wikidata has helped make their data and knowledge public, which would have otherwise remained in their private library systems and been inaccessible to others. As a member of the cataloging team, she recognized that the amount of data her team created was often invisible to the public, and Wikidata provided a way for the team to share their work output outside of their organization:

> *We make a lot of records and stuff, but they stay in library systems, so cataloging sometimes feels like people don't understand the amount of knowledge that comes through here, and that maybe we don't get to share it as much. So, by participating in Wikidata, it means that all the info that they're also putting into these LC name authority records can get outside of a library.*

*5.1.2 Harnessing the Wisdom of Wikidata Crowds.* Harnessing the wisdom of Wikidata crowds is a motivation for providers, acquirers, and mutualists. As of October 2024, Wikidata has nearly 24,000 active users [7]. Some of our participants noted the value of Wikidata contributors' participation in improving their records. For example, P3 has found that Wikidata contributors are particularly helpful in completing or filling in unknown fields of new entities: *"I think the beauty of Wikidata is that so many people participate so that we don't have to build records for all of these authors. I just have to make sure that they're there, and then I can query them"*. P10's organization is also motivated by the potential of involving a wide range of contributors in their data work. As P10's specific focus is software preservation and emulation, a very niche area, she sees outreach to Wikidata contributors as particularly beneficial:

> *We want to have as many people as possible doing this work because it's so wide-ranging as well as being so specific[...] We just wanted to get as many people as we could involved in this project. And so Wikidata sort of felt like a natural path to go down[...] There's value in information coming from as many different people as possible and not having that siloed to one institution or just a few institutions.*

Acquirers are particularly interested in leveraging Wikidata contributors' knowledge to fill in gaps in the information about their collections. P12 valued the information from Wikidata because, without it, many objects in their access system would have limited information for visitors, and *"it would require a lot of human resources to complete data, to fill up databases, fields, and so on"*. Similarly, P13 and P14, who work together at a private archive, rely on Wikidata as a source to add birth dates to entities in their digital collections. Such metadata is essential for cataloging records of people, such as authors with common names.

*5.1.3 A Hub of Identifiers.* Wikidata's unique feature as an identifier hub is a strong motivation for all archetypes. On Wikidata, each entity is identified with a unique Wikidata identifier and also has other identifiers used by national libraries, prominent collection systems, and individual institutions, such as ISBN-13. The abundance of identifiers currently on Wikidata makes it possible for all types of memory institutions to link managed entities correctly with Wikidata entities when uploading or downloading data. Memory institutions can also add their institutions' identifiers to Wikidata by proposing their identifier as a new property [6]. Three organizations (two museums and one archive, i.e., P4's, P12's, and P8 and P15's) from our pool have successfully integrated their organizational identifiers into the Wikidata entities related to their collections. P2, an expert cataloger and community organizer, has been working with other professionals to examine and refine entity management in libraries. For her, Wikidata's unique feature as an identifier hub is a resource for libraries:

> *[Our group is] trying to look at the way that libraries are managing linked data entities. Oh, this thing in Wikidata is the same as or different from this thing in NACO, which is the same or different from this thing in the National Library of Greece or [...] Because that's the whole point of linked data and the whole point of what we're trying to do in entity management and Wikidata is really good for that.*

*5.1.4 Support for Discovery.* Using Wikidata to support discovery services and systems is another motivation for providers and mutualists. As Wikidata is a shared data infrastructure, information professionals can aggregate and query all the entities on Wikidata, not just the ones in their collections. Motivated by this capability, P6's organization built a collection system that pulls, connects, and aggregates data from Wikidata and other sources. P6 envisioned that if other institutions adopt the same collection system, together they can build a virtual "One World Museum" that leverages collections hosted by member institutions and enables users to search through a larger, unified knowledge base.

P3, a provider, also values Wikidata's support for discovery. She searches the database regularly to gain a better understanding of data related to the organization's collections. While her organization does not download data from Wikidata, she frequently uses the SPARQL query [3] service to aggregate data connected to the organization to gain insights into the organization's collections. For example, she queried all Wikidata entities whose "archive at" fields include their organization, aggregating the educational background of these entities to understand where these people received their degrees. Even though P3 did not contribute any information about these people's educational backgrounds to Wikidata, she was able to retrieve that information. As P3 put it, "*You didn't have to create it; you just had to create that link.*"

*5.1.5 Storage and Synchronization.* Storage is a motivation for providers and mutualists who have data or information that they do not have a place to store. When the local collection system that hosts the organization's theses and dissertations was being retired, P1 turned to Wikidata as a storage place to save the metadata. P7 mentioned that her team uses Wikidata so they can keep track of entities that do not have a VIAF (Virtual International Authority File) or Library of Congress ID:

> *Often our work as meta librarians might involve researching who a person was, and we don't have a place to put that information. And Wikidata is a really great place. For instance, if you're researching the subject of a photograph and you find all this biographical information, you can put it into Wikidata.*

P5 and P9 also turned to Wikidata as a storage space for the metadata from their collections. In particular, they valued Wikidata's flexible requirements for data completeness: they can always enrich their entities as they uncover new information about them, and do not have to fill out many fields at once, as long as the entity meets Wikidata's notability [4] and quality requirements.

> *The requirements we got from [another data management system] were a little bit stricter in terms of wanting specific kinds of information that we didn't always have for everyone. So we started looking at and exploring Wiki data as a possibility and a way for us to begin to think about how to model the data and contribute those names.*

Relatedly, Wikidata as a storage space allows mutualists to synchronize their organization's collective work, especially when working in large teams or collaborating with multiple units. P6 contributes to Wikidata, in part because it will enable other units in his organization to receive the most up-to-date data:

> *We're updating that information not in the collections management system for the Museum or in the collections management system for the art gallery. We're doing it in Wikidata and then reconciling information from each local database to the external Wikidata identifiers. That way, you get the same information in all places.*

### 5.2 Practices and Challenges

The institutions in our interview pool have different practices and processes when using Wikidata. This comes with varied collaboration approaches and challenges. Below, we describe how providers and mutualists contribute diverse metadata, collaborate within their institutions, and the challenges they run into when contributing to and querying Wikidata.

---

[3] https://en.wikipedia.org/wiki/SPARQL
[4] https://www.wikidata.org/wiki/Wikidata:Notability

*5.2.1 Diverse Types of Contributions.* Providers and mutualists like universities and public libraries contribute various types of data to Wikidata, ranging from faculty and researcher profiles to research outputs (books, monographs, publications, and dissertations) to special collections. For example, P3 mentioned two specific areas of contributions: creating entities for university-affiliated researchers and augmenting existing Wikidata entities related to specific artifacts at their organization. This work is motivated by P3's institutional goal of increasing the visibility of the institution's research impact. Similarly, P1 mentioned how the institutional goal shaped his Wikidata contributions: *We're raising the profile of our research and scholarship because that is something [university leadership is] really interested in."*

Some participants manage subject-specific collections like natural history specimens and are focused on enriching Wikidata records related to specific subjects or artifacts. P5, who works in an architecture library under a university library system, mainly contributes metadata on buildings. Similarly, P10 focuses on contributing software metadata to Wikidata to give more context to people accessing their software emulation collections.

Another unique type of contribution is metadata on file formats. P8 and P15 are from a government archive, and their team specializes in risk assessment for file formats and preservation planning. They contributed metadata on file formats to Wikidata. P8 saw Wikidata as a definitive reference guide for resources related to the digital preservation of obsolete formats.

> *[We are] trying to use Wikidata more as a centralized hub for where all of these different resources about digital preservation and about file formats can all point to.*

Some participants took a more open-ended approach when deciding what contributions to make. P3's team considers expertise, perceived benefit to users, and potential impact when deciding what to contribute to Wikidata: *"We sort of do whatever we want: whatever intrigues us and whatever we think is going to benefit our users and our data in the future. [We do] whatever we think is achievable and whatever we think is going to be best for the profession."*

*5.2.2 Collaboration Processes (or Lack Thereof).* We observed different collaboration processes and noticed that larger institutions with more staff members tend to have the ability to afford experts to work on various areas of contribution. P2 and P3, who work in large academic libraries of public universities, have catalogers with different areas of expertise who run projects and contributions based on their areas of knowledge and interests. Some may also consult and collaborate with faculty members and scholars who study a specific subject area. Some providers and mutualists organized well-scoped, one-time projects with students in training. Both P3 and P6 have worked with or trained students to create Wikidata entities for specific collections.

Notably, despite Wikidata's potential for collaborative work, the Wikidata PCC Pilot Project was the only instance in which multiple institutions worked together[8]. Through the project, participants collaborated with other member institutions, such as planning onboarding, developing reusable workflows, and designing shared best practices. Although we saw little evidence of information professionals collaborating across institutional boundaries, they expressed a strong interest in working together. Both P8 and P10 noted that Wikidata could serve as a centralized hub to share information; however, due to hour constraints and shifting priorities, they did not explore collaboration opportunities with other institutions.

For some participants, contributing to Wikidata is a solitary journey. P1 mentions that there is no coordinated university-wide initiative for Wikidata. Many university employees contribute at an individual level: *"We're given some freedom in our scholarship and service, so individual people* *will take it on.*" Even though P11 started contributing to Wikidata as part of a team, she said she is the only one active currently.

*5.2.3 Challenges and Barriers.* All interviewees reported barriers to entry and a steep learning curve, both in terms of contributing and querying. OpenRefine and QuickStatements are frequently mentioned by participants who are expert contributors, but some participants who are not part of a learning community or network expressed uncertainty about where to start. Many expert contributors who actively query and contribute to Wikidata come from an academic background in library and information sciences and are involved in the LD4 community–a network of information professionals interested in linked data. They acknowledge that their background and daily work of dealing with large databases have helped them use SPARQL (Wikidata's native query service) and OpenRefine with ease.

Knowing when to stop is another common challenge faced by Wikidata contributors. Due to the nature of linked data, participants find it easy to fall into the trap of going down rabbit holes in creating records. While some, like P3, find it an exhilarating endeavor, P9 mentioned how it can lead to creating records about entities on which they may not have enough information, compromising the robustness of those records:

> *If you're not careful, you can end up in a place where you're like, 'I'm trying to create this thing for an award that I know nothing about.' Then we shouldn't be creating that. We're not the people who have that particular level of expertise or knowledge.*

Some participants also expressed conflicts with other Wikidata contributors due to different expectations of standards, particularly those concerned with data about marginalized groups. P2 talked about her frequent encounters with an administrator on Wikidata who added unverified gender information to the records of people from their university:

> *There are a lot of people who add gender information to our people, and it's really annoying. I went in and undid a couple of revisions last week by this person who's an admin in Wikidata, and I have [asked her] to stop adding binary genders based on their first names or based on pronouns and she just doesn't respond to me. And she's doing it constantly to people in our community. And there's no recourse really.*

Providers and mutualists also found monitoring edits or revsions on entities they created challenging. Currently, the creator of an entity will receive a notification automatically when new edits are made. However, as one creates more items, they will receive more notifications around edits, which can quickly become time-consuming to read through. P2, who has contributed and created thousands of items, finds difficult to keep track of changes, especially when there are conflicting or unverified edits around gender identities and personal pronouns.

Participants mentioned a lack of community guidelines as a barrier to contribution. Novice providers reported instances of contributions being changed or deleted by other contributors or bots, either due to non-compliance with Wikidata's notability requirements or conflicting views. While these participants expressed understanding and acceptance of such changes, they also expressed their desire to learn more about the standards and notability requirements when contributing to Wikidata to avoid such instances in the future. P2 also mentioned that as the community is open and does not set governance guidelines, differences in perspectives make collaboration difficult.

Some participants expressed that current Wikidata schemas do not support the contribution of artifacts from indigenous communities. P4 expressed concerns about the relationship between Wikidata and indigenous knowledge: *"The basic idea of universal access to all knowledge, the norms, the expectations, and the way the rules are set up are not built well to accommodate indigenous* *knowledge "*. She also mentioned how particular objects and artifacts of cultural importance and history could be mistranslated when contributed to Wikidata and lack local contexts and personal, traditional, and spiritual histories. Some institutions, like the one of P12, have designated roles that create standards and guidelines when dealing with indigenous data, but still find it challenging to contribute these artifacts to Wikidata.

Providers, mutualists, and acquirers all expressed constraints around resources when working with Wikidata. Some participants, like P10, have been reconsidering the sustainability of their approach: *"We need to take a step back and try to figure out if we need to build additional tooling or documentation to bring more people into the work rather than having it be three people do all of this work"*. Some institutions often experiment with different approaches by delivering well-scoped, one-time projects, such as an edit-a-thon or hiring student assistants to focus on a specific collection. P13 mentions that her team started a one-time project while out of the office during COVID-19 to link Wikidata entities with subjects in their archives because such tasks can be done remotely. P12's team collaborated on a one-off project with a Wikimedian-in-Residence to extract information from Wikidata and display it on their website along with Wikidata links: *"There was a grant from Wikimedia. The project has now been delivered, and unfortunately, there wasn't any follow-up because of a lack of resources."* These experimental approaches help institutions explore working with Wikidata under their resource constraints.

### 5.3 Wikidata as a Tracking Apparatus

Wikidata makes every editor's past contributions open source, so it is possible to trace the provenance of every piece of information. Notably, we did not notice any participant using this feature to gain credits for their contributions, unlike the case with Wikipedia editors [25]. Rather, providers and mutualists rely on this feature to advocate for resources, support accountability, gain insights into their collective progress, and gauge their impact.

*5.3.1 Advocating for Resources.* Most participants did not actively keep track of their Wikidata contributions and were uncertain whether this would interest their supervisor; however, some tracked their numbers to report to supervisors and advocated for resources. P1 reported they kept track of items added to Wikidata for annual reports. Relatedly, P7 expressed interest in better tooling to track their work, like: *"If I could, for instance, do a query [directly in Wikidata] and say 'show me all the people archived at our library and show me the ones that I created on Wikidata', so I could tell my boss, that would be useful"*. P8's team ensured supervisors were aware of their significant contributions: *"When I did the first big upload with quick statements of over 500 identifiers and [we went] from having nothing of ours in Wikidata to having over 500. That was something we made sure that our office executive was aware of. It was a big deal that we had gotten that much data in."*

*5.3.2 Supporting Accountability.* Some participants also relied on Wikidata's edit history to hold themselves accountable. P2, whose Wikidata profile page stated her organizational affiliation, always logged in before contributing to maintain accountability. P8 emphasized transparency over recognition, especially given her novice Wikidata contributor status: *"I like having my name on it, so it's like if I was making some error a lot, someone who actually knows me could be like, hey, you keep doing this thing, stop doing that."*

*5.3.3 Providing Insights into Collective Progress.* Although tracking one's own contributions on Wikidata was uncommon, many participants reported monitoring collective organizational progress through various tools. The Outreach Dashboard hosted by Wiki Education aggregates team contributions, although some noted accuracy limitations with project-irrelevant contributions. P2's team employs Listeria for project-specific contribution tracking, focusing on collective rather than individual metrics. P2 stressed that she does not track her individual contributions, but tracks their collective progress for specific projects more closely. Their purpose for tracking is *"a lot more project-oriented than individual-oriented"*. Other approaches included P4's development of Python scripts to analyze the team's contribution history and to guide future priorities.

*5.3.4 Tracking Impact Within and Beyond Wikidata.* Participants sought to understand both the immediate and downstream impacts of their contributions. Within Wikidata, some actively monitor changes to their entities through watchlists and queries. For example, P1 regularly checked how volunteers and bots enhanced the entities he created. In a similar vein, P5 used Wikidata's query services to assess improvements to team contributions and found that even minor bot-added details can enhance record discoverability.

However, tracking impact beyond Wikidata proved challenging for our participants. While participants are aware of Wikidata's potential impact as an open linked data infrastructure, many expressed uncertainty about the lack of mechanisms to measure the reach of contributions. Some relied on anecdotal evidence, such as P7's observation about increasing collection visibility: *"adding an 'archives at' statement to a person's Wikidata page means that when you search that person, it helps search engine surface the finding aid from your collection"*. P8 also noted that the lack of a window into their contributions' broader impact outside of Wikidata has hindered our participants' ability to advocate for more resources. Our participants desired tools to see how their contributions may increase public engagement with their collections in the form of web traffic, in-person visits, presence in search engines, and scholarly citations.

## 6 Discussion

In this section, we begin by discussing opportunities to leverage information professionals' training and background to improve contributions, encourage collaboration, and better understand the value of the data work being conducted. We then explore ways to transform providers and acquirers into mutualists and unpack the design, policy, and education implications of our findings about information professionals' practices and challenges.

### 6.1 Supporting Collaborative Work on Wikidata

*6.1.1 Between Information Professionals and Volunteers.* Our study highlights the distinct motivations and practices of information professionals compared with volunteer contributors on Wikidata. In particular, providers and mutualists are often driven by their institutional goals of open access and engagement, unlike volunteer contributors who are driven by personal interests in structured data [71]. Additionally, information professionals handle a large amount of data from their collections, and some rely heavily on automation tools such as OpenRefine and QuickStatements to make changes and manage records. This practice sets them apart from the kind of volunteer contributors that Zhang et al. call "Masons"–volunteer contributors who perform individual, easy-to-complete tasks of data entry and editing [71]. Last but not least, information professionals may have different standards than volunteer contributors, stemming from their professional training to ensure an accurate representation of people and indigenous knowledge and artifacts [26]. These differences point to opportunities to support the collaborative, complementary work between information professionals and volunteers.

Information professionals' contributions may serve as a scaffold to organize volunteer efforts. As our participants pointed out, additions made by volunteer contributors help to enrich the entities they have created. As such, it may be helpful for memory institutions to start building communities of volunteer contributors around the contributions they make. This could take the form of a Wiki

Project in which volunteers and information professionals work together to create and update entities related to a memory institution's holdings. Strengthening this collaboration will also improve the sustainability of peer production on Wikidata; volunteer participation will help ensure that this peer production project is not unduly affected by the priorities of organizations that sponsor contributions [66, 73]. Volunteers will benefit from existing norms, guidelines, and institutional knowledge to continue their Wiki Projects. Some acquirers reported that they haven't been able to contribute to Wikidata due to resource constraints and shifting priorities. In collaboration with the Wikimedia Foundation, memory institutions can host more training programs and workshops for volunteers to help extend the reach of their existing collections. These actions would support the training and development of volunteers and their staff collectively, and encourage further collaboration opportunities with volunteers and other memory institutions. These initiatives will also create opportunities for volunteers, information professionals, and Wikimedia to collectively shape the future directions of Wikidata, along with helping memory institutions' progress towards their open access goals.

Our participants reported some experiences with conflicting edits or views on Wikidata with other contributors, pointing to an opportunity to facilitate more productive conversations around such conflicts. Our participants' expectations and perspectives are shaped by their professional training, particularly around accurately representing people, such as gender information [31, 68] and indigenous knowledge. Information professionals may benefit from tools that can monitor frequently contested contributions or conflicted changes to so that they can easily start a discussion about their contributions and rationales with the editors of such entities. Such tools may even include template messages for information professionals to use, given the large number of entities they tend to create in batches. Correspondingly, Wikidata could also put up an approval process for edits of frequently conflicted entities. For example, an edit for the gender property could require two approvals before the change is published. This can help facilitate subtle collaboration and discussions between volunteers and information professionals, especially on contested entities. Similarly, Wikidata's notability requirement may also benefit from perspectives from information professionals, who could leverage their expertise to verify significant figures backed with references.

*6.1.2 Between Memory Institutions.* Although Wikidata is well positioned to support collaboration between institutions, we saw very limited evidence of information professionals collaborating with each other, except for the Wikidata PCC pilot project. This might be due to their focus on specific collections in their holdings.

Collaboration among memory institutions has the potential to enrich Wikidata as an open-linked data infrastructure and downstream knowledge technologies (e.g., Wikipedia and search engines). In fact, our participants see the value of collaboration through Wikidata. As such, we see a valuable opportunity for memory institutions to pair up with those with similar needs or related collections. Existing events and gatherings attended by information professionals (Wikimedia-affiliated and beyond) could serve as promising venues for information professionals to share and exchange skills beyond their institutional boundaries while helping to enrich Wikidata. This may take the form of interactive workshops organized by information professionals, oriented toward different groups such as providers, acquirers, and mutualists.

### 6.2 The value of Information Professionals' Data Work

Our study contributes to the literature on data work by highlighting the practices of information professionals in creating new data on Wikidata, updating it, and managing their records (all forms of data work) [40, 49]. Information professionals bring value to Wikidata through their expertise and training in implementing metadata standards, authority control, and ethical considerations.

Along with their expertise, these professionals are contributing valuable data to Wikidata and/or their organizations' collections. Their data work creates significant value for Wikidata, sibling projects like Wikipedia, and memory institutions' capacities to build open access points consistent with prior work on data sharing [65]. Providers and mutualists systematically contribute specialized data from their collections, including architectural records, software metadata, and file format specifications. These contributions, crafted by information professionals with subject area expertise, can increase the accuracy and breadth of Wikidata topics while advancing institutional open access goals. As P3 explained, the quality, depth, and quantity of records created and stewarded by memory institutions often remain invisible in library management systems. Wikidata provides a venue for making this valuable data work publicly accessible and reusable.

As part of their data work with Wikidata, information professionals undertake stewardship activities that are distinct from typical volunteer contributors. For example, some of our participants actively monitor changes to their contributions, hold editors accountable for accuracy, and adhere to the professional standards of librarians and archivists working in research institutions [4][1]. These activities support the maintenance of Wikidata and ensure the integrity of data [15, 35, 55]. When information professionals invest in stewardship activities like this, they bring both Wikidata infrastructure and memory institutions into further alignment for both preservation and access.

Our work highlights the importance of using a wider lens to examine data work and consider its broader impact. Like many other types of open data, information professionals' data work benefits many downstream technologies like search engines, conversational agents, LLMs, and other "invisible machines" that consume open knowledge [71]. With more and more open data being generated and shared, understanding data work's impact and value requires an understanding of "the context of production" of data [65]. Consistent with the "Collections as Data" movement mentioned above [45], our provider and mutualist participants believe in making entities and records free and public to broaden the impact and access of their digital archives and repositories. Their data work puts these beliefs into action to enrich Wikidata as a valuable infrastructure for all. As such, we argue for considering the value of data work holistically, not just within individual systems, but for the broader knowledge infrastructure.

### 6.3 Fulfilling Wikidata's Potential–Roundtrips

Uploading and exchanging data with Wikidata, as mutualists do, is also known as creating a "round trip" [23]. A research report by Wikimedia Deutschland e.V[5] interviewed members of cultural institutions and identified strong interests in creating round trips between their databases and Wikidata. However, in our study, we only identified three mutualist organizations that created data round trips. Others focus solely on uploading or downloading data and thereby miss opportunities to leverage Wikidata to either improve their local records or enrich Wikidata with their valuable data. Below, we discuss how providers and acquirers might experiment with Wikidata to become mutualists by leveraging the existing mapping between their own collections and Wikidata entities.

*6.3.1 Providers to Mutualists.* Providers with their own collection systems may benefit immediately from querying Wikidata regularly to update their local databases. Providers may first focus on extracting data from prominent fields that are known to contain robust information. Acquirers P13 and P14 did this by carefully selecting the Wikidata fields to enrich their collections. Providers may also consider introducing manual verification on a subset of entities in their collections before updating their databases at scale. This would allow providers to understand how other Wikidata contributors have edited the entities in their collections before systematically updating their records based on Wikidata.

---

[5]https://meta.wikimedia.org/wiki/Wikimedia_Deutschland

*6.3.2 Acquirers to Mutualists.* All acquirers from our sample expressed a strong interest in contributing back to Wikidata, but did not due to expertise and resource constraints. Given these constraints, acquirers may benefit from establishing stronger connections with Wikidata communities and peer institutions with extensive expertise and tools who are both providers and mutualists, such as participants of the Wikidata PCC pilot project [6]. Peer institutions' experiences with automated tools such as OpenRefine may help to minimize the number of staff hours required by acquirers to upload their local collections to Wikidata. It may be particularly helpful to connect the acquirers with peer institutions that have a strong commitment to openness. The latter's expertise in connecting Wikidata with access infrastructures may help to guide acquirers to better prioritize what types of information to contribute back to Wikidata.

### 6.4 Tools, Policies, and Training for Information Professionals

*6.4.1 Tools and Metrics for Tracking.* Prior investigation has found that information professionals need metrics to understand the impact of their contributions [23]. Our participants reported aggregating and tracking their contributions as a way to understand their impact, but also faced challenges in understanding how their Wikidata contributions benefit knowledge infrastructures and downstream technologies. This potentially limits their ability to effectively organize their efforts and fulfill their mission of sharing knowledge and broadening access. Researchers and developers who are interested in supporting memory institutions and Wikidata communities may explore developing tools and metrics that can assess information professionals' impact on improving access and addressing knowledge gaps. For example, given Wikidata's prominent role in supporting knowledge graphs and model training (e.g., [39, 60]), developers may examine the utility of Wikidata contributions to AI models, particularly those related to rare and special collections in memory institutions, and make transparent the value of these contributions.

*6.4.2 Institutional Policies for Data Management.* The wide range of contributions made by information professionals illustrates varied and extra-institutional work that would benefit from more data management policies, either from their own memory institutions or from Wikidata itself. For providers and mutualists, capturing the documentation and tracking of Wikidata contributions is crucial to demonstrate impact and merit institutional support. For mutualists and acquirers who integrate Wikidata into their collections, a data provenance policy that supports systematic verification would bolster trust and ensure consistency with institutional standards. Standardized metadata documentation in Wikidata would benefit all memory institutions' data workers, allowing for increased consistency between memory institutions' collections systems and open platforms.

*6.4.3 Training for Information Professionals.* All participants recognized the sharp learning curve in contributing to Wikidata and identified various challenges even after becoming expert Wikidata contributors. These challenges highlight the need to integrate Wikidata into the formal curriculum for training information professionals and to provide continuous education opportunities. For example, educators may integrate Wikidata into formal MSIS/MLIS/archival studies curricula, covering both theoretical foundations and hands-on practice with open platforms. These programs may also include targeted technical skills and training modules focusing on SQL and query languages, API integration, and data exchange protocols. Information professional communities may develop continuing education programs that leverage existing Wiki resources, such as the Outreach dashboard and experienced Wikimedian, to provide ongoing support and mentorship for professionals at different skill levels. Furthermore, the Wikimedia Foundation could also leverage existing networks of information professionals like the LD4 community to encourage institutions to join the Wikidata movement by hosting onboarding workshops and tutorials.

---

[6]https://www.wikidata.org/wiki/Wikidata:WikiProject_PCC_Wikidata_Pilot

Acquirers and mutualists rely on OpenRefine to reconcile and link entities between their collections and Wikidata. This process can be laborious. It may be beneficial to further understand what challenges information professionals face when they use OpenRefine to synchronize their records and Wikidata, enhancing their own records, and metadata enrichment. It may be particularly valuable to develop other automated tools and train information professionals to support seamless data synchronization.

### 6.5 Limitations and Future Work

As a qualitative interview study, this work focused on information professionals' self-reported motivations, practices, and processes and may overlook activities that they perform regularly on Wikidata but did not describe in our interviews. Additionally, given our sample size, our findings should not be generalized to all information professionals who work with Wikidata. For example, we cannot make any conclusions about the percentages of providers, acquirers, and mutualists in memory institutions. Future work may leverage surveys and trace data to gain a more structured understanding of information professionals' data work with Wikidata. Moreover, given that collections management varies from institution to institution, our recommendations may not be suitable for some.

## 7 Conclusion

We studied why and how information professionals from memory institutions use Wikidata for data work. We identified three archetypes of Wikidata users– providers, acquirers, and mutualists– as well as various types of contributions information professionals make and their processes. Despite Wikidata's potential for supporting collaborative work, we saw limited evidence of information professionals collaborating across institutional boundaries. While many information professionals monitor their collective Wikidata contributions, the impact on downstream technologies is often unclear to them. We discuss opportunities for information professionals and Wikidata stakeholders to support collaborative work across institutions and with volunteer contributors. We discuss the implications of information professionals' data work on the broader knowledge ecosystem, as well as opportunities to leverage Wikidata to create data round trips. We also provide design, policy, and training recommendations to bolster memory institutions' integration with Wikidata.

## References

[1] n.d. ALA Code of Ethics | ALA. https://www.ala.org/tools/ethics

[2] n.d. Code of Ethics for Museums. https://www.museumsassociation.org/campaigns/ethics/code-of-ethics/

[3] n.d. Information Maintenance as a Practice of Care: An Invitation to Reflect and Share | Maintainers. https://themaintainers.org/information-maintenance-as-a-practice-of-care-an-invitation-to-reflect-and-share/

[4] n.d. SAA Core Values Statement and Code of Ethics | Society of American Archivists. https://www2.archivists.org/statements/saa-core-values-statement-and-code-of-ethics

[5] n.d. Wikidata:Bots - Wikidata. https://www.wikidata.org/wiki/Wikidata:Bots

[6] n.d. Wikidata:Property proposal - Wikidata. https://m.wikidata.org/wiki/Wikidata:Property_proposal

[7] n.d. Wikidata:Statistics - Wikidata. https://www.wikidata.org/wiki/Wikidata:Statistics

[8] n.d. Wikidata:WikiProject PCC Wikidata Pilot - Wikidata. https://www.wikidata.org/wiki/Wikidata:WikiProject_PCC_Wikidata_Pilot

[9] Amelia Acker. 2021. Metadata. In *Uncertain Archives: Critical Keywords for Big Data*, Nanna Thylstrup, D. Agostinho, A. Ring, C. D'Ignazio, and K. Veel (Eds.). MIT Press, 321–330. https://direct.mit.edu/books/edited-volume/5002/chapter/2654158/Metadata

[10] Tarfah Alrashed, Lea Verou, and David Karger. 2022. Wikxhibit: Using HTML and Wikidata to Author Applications that Link Data Across the Web. In *Proceedings of the 35th Annual ACM Symposium on User Interface Software and*

*Technology*. ACM, Bend OR USA, 1–15. doi:10.1145/3526113.3545706

[11] Gabriel Amaral, Alessandro Piscopo, Lucie-aimée Kaffee, Odinaldo Rodrigues, and Elena Simperl. 2021. Assessing the Quality of Sources in Wikidata Across Languages: A Hybrid Approach. *J. Data and Information Quality* 13, 4 (Oct. 2021), 23:1–23:35. doi:10.1145/3484828

[12] Patti Bao, Brent Hecht, Samuel Carton, Mahmood Quaderi, Michael Horn, and Darren Gergle. 2012. Omnipedia: bridging the wikipedia language gap. In *Proceedings of the SIGCHI Conference on Human Factors in Computing Systems (CHI '12)*. Association for Computing Machinery, New York, NY, USA, 1075–1084. doi:10.1145/2207676.2208553

[13] Karthik S Bhat, Mohit Jain, and Neha Kumar. 2021. Infrastructuring Telehealth in (In)Formal Patient-Doctor Contexts. *Proceedings of the ACM on Human-Computer Interaction* 5, CSCW2 (Oct. 2021), 1–28. doi:10.1145/3476064

[14] Christine L. Borgman. 2017. *Big Data, Little Data, No Data: Scholarship in the Networked World* (reprint edition ed.). The MIT Press, Cambridge, Massachusetts London, England.

[15] Claus Bossen and Kathleen H. Pine. 2023. Batman and Robin in Healthcare Knowledge Work: Human-AI Collaboration by Clinical Documentation Integrity Specialists. *ACM Transactions on Computer-Human Interaction* 30, 2 (April 2023), 1–29. doi:10.1145/3569892

[16] Claus Bossen, Kathleen H Pine, Federico Cabitza, Gunnar Ellingsen, and Enrico Maria Piras. 2019. Data work in healthcare: An Introduction. *Health Informatics Journal* 25, 3 (Sept. 2019), 465–474. doi:10.1177/1460458219864730 Publisher: SAGE Publications Ltd.

[17] Cara Bradley. 2021. Academic Librarians, Open Access, and the Ethics of Care. *Journal of Librarianship and Scholarly Communication* 9, 1 (July 2021). doi:10.31274/jlsc.12914 Number: 1 Publisher: Iowa State University Digital Press.

[18] Susan L. Bryant, Andrea Forte, and Amy Bruckman. 2005. Becoming Wikipedian: transformation of participation in a collaborative online encyclopedia. In *Proceedings of the 2005 international ACM SIGGROUP conference on Supporting group work - GROUP '05*. ACM Press, Sanibel Island, Florida, USA, 1. doi:10.1145/1099203.1099205

[19] Gustavo Candela, Sally Chambers, and Tim Sherratt. 2023. An approach to assess the quality of Jupyter projects published by GLAM institutions. *Journal of the Association for Information Science and Technology* 74, 13 (2023), 1550–1564. doi:10.1002/asi.24835 _eprint: https://onlinelibrary.wiley.com/doi/pdf/10.1002/asi.24835.

[20] Gustavo Candela, María Dolores Sáez, MPilar Escobar Esteban, and Manuel Marco-Such. 2022. Reusing digital collections from GLAM institutions. *Journal of Information Science* 48, 2 (April 2022), 251–267. doi:10.1177/0165551520950246 Publisher: SAGE Publications Ltd.

[21] Giovanni Colavizza, Tobias Blanke, Charles Jeurgens, and Julia Noordegraaf. 2021. Archives and AI: An Overview of Current Debates and Future Perspectives. *J. Comput. Cult. Herit.* 15, 1 (Dec. 2021), 4:1–4:15. doi:10.1145/3479010

[22] Erenrich. 2024. Wikidata & AI, together again. https://tech-news.wikimedia.de/en/2024/02/19/wikidata-ai-together-again/

[23] Jan Dittrich e.V, Wikimedia Deutschland. 2019. English: We talked to 16 users who worked at different cultural ("GLAM") institutions to find out about "How and why do people in cultural institutions use Wikidata?" and thus learn more about participants' motivations, activities and problems. We did the research from June 2019 - September 2019. https://commons.wikimedia.org/wiki/File:Research_Report_%E2%80%93_Use_of_Wikidata_in_GLAM_institutions_(2019-11).pdf

[24] Shani Evenstein Sigalov and Rafi Nachmias. 2023. Investigating the potential of the semantic web for education: Exploring Wikidata as a learning platform. *Education and Information Technologies* 28, 10 (Oct. 2023), 12565–12614. doi:10.1007/s10639-023-11664-1

[25] Andrea Forte and Amy Bruckman. n.d. Why Do People Write for Wikipedia? Incentives to Contribute to Open-Content Publishing. (n.d.).

[26] Richard Gartner. 2016. *Metadata: Shaping Knowledge from Antiquity to the Semantic Web*. Springer. Google-Books-ID: 2Z_VDAAAQBAJ.

[27] Andrew Hall, Loren Terveen, and Aaron Halfaker. 2018. Bot Detection in Wikidata Using Behavioral and Other Informal Cues. *Proceedings of the ACM on Human-Computer Interaction* 2, CSCW (Nov. 2018), 1–18. doi:10.1145/3274333

[28] Azra Ismail and Neha Kumar. 2018. Engaging Solidarity in Data Collection Practices for Community Health. *Proceedings of the ACM on Human-Computer Interaction* 2, CSCW (Nov. 2018), 1–24. doi:10.1145/3274345

[29] Lucie-Aimée Kaffee, Kemele M Endris, and Elena Simperl. 2019. When Humans and Machines Collaborate: Cross-lingual Label Editing in Wikidata. (2019).

[30] Lucie-Aimée Kaffee, Alessandro Piscopo, Pavlos Vougiouklis, Elena Simperl, Leslie Carr, and Lydia Pintscher. 2017. A Glimpse into Babel: An Analysis of Multilinguality in Wikidata. In *Proceedings of the 13th International Symposium on Open Collaboration*. ACM, Galway Ireland, 1–5. doi:10.1145/3125433.3125465

[31] Brian C. Keegan and Jed R. Brubaker. 2015. 'Is' to 'Was': Coordination and Commemoration in Posthumous Activity on Wikipedia Biographies. In *Proceedings of the 18th ACM Conference on Computer Supported Cooperative Work & Social Computing (CSCW '15)*. Association for Computing Machinery, New York, NY, USA, 533–546. doi:10.1145/2675133.2675238

[32] Rob Kitchin. 2014. *The Data Revolution: Big Data, Open Data, Data Infrastructures &amp; Their Consequences.* SAGE Publications Ltd. doi:10.4135/9781473909472

[33] Aniket Kittur and Robert E. Kraut. 2008. Harnessing the wisdom of crowds in wikipedia: quality through coordination. In *Proceedings of the 2008 ACM conference on Computer supported cooperative work (CSCW '08)*. Association for Computing Machinery, New York, NY, USA, 37–46. doi:10.1145/1460563.1460572

[34] Karim R. Lakhani and Robert G. Wolf. n.d. Why Hackers Do What They Do: Understanding Motivation and Effort in Free/Open Source Software Projects. (n.d). https://direct.mit.edu/books/edited-volume/3816/chapter/125228/Why-Hackers-Do-What-They-Do-Understanding

[35] The Information Maintainers, D. Olson, J. Meyerson, M. A. Parsons, J. Castro, M. Lassere, D. J. Wright, H. Arnold, A. S. Galvan, P Hswe, B. Nowviskie, A. Russell, L. Vinsel, and A. Acker. 2019. Information Maintenance as a Practice of Care. (June 2019). doi:10.5281/zenodo.3251131 Publisher: Zenodo.

[36] Matthew S Mayernik. 2019. Metadata accounts: Achieving data and evidence in scientific research. *Social Studies of Science* 49, 5 (Oct. 2019), 732–757. doi:10.1177/0306312719863494 Publisher: SAGE Publications Ltd.

[37] Connor McMahon, Isaac Johnson, and Brent Hecht. 2017. The Substantial Interdependence of Wikipedia and Google: A Case Study on the Relationship Between Peer Production Communities and Information Technologies. *Proceedings of the International AAAI Conference on Web and Social Media* 11, 1 (May 2017), 142–151. doi:10.1609/icwsm.v11i1.14883

[38] Florence Millerand. 2012. Network Science,"Invisible" Information Managers in the Production of a Scientific Database. *Revue d'Anthropologie des Connaissances* 6, 1, 1 (2012), 163. doi:10.3917/rac.015.0201

[39] Fedor Moiseev, Zhe Dong, Enrique Alfonseca, and Martin Jaggi. 2022. SKILL: Structured Knowledge Infusion for Large Language Models. doi:10.48550/arXiv.2205.08184 arXiv:2205.08184 [cs].

[40] Naja Holten Møller, Claus Bossen, Kathleen H. Pine, Trine Rask Nielsen, and Gina Neff. 2020. Who does the work of data? *interactions* 27, 3 (April 2020), 52–55. doi:10.1145/3386389

[41] Claudia Müller-Birn, Benjamin Karran, Janette Lehmann, and Markus Luczak-Rösch. 2015. Peer-production system or collaborative ontology engineering effort: what is Wikidata?. In *Proceedings of the 11th International Symposium on Open Collaboration (OpenSym '15)*. Association for Computing Machinery, New York, NY, USA, 1–10. doi:10.1145/2788993.2789836

[42] Andrew B. Neang, Will Sutherland, Michael W. Beach, and Charlotte P. Lee. 2021. Data Integration as Coordination: The Articulation of Data Work in an Ocean Science Collaboration. *Proceedings of the ACM on Human-Computer Interaction* 4, CSCW3 (Jan. 2021), 1–25. doi:10.1145/3432955

[43] Oded Nov. 2007. What motivates Wikipedians? *Commun. ACM* 50, 11 (Nov. 2007), 60–64. doi:10.1145/1297797.1297798

[44] Thomas Padilla, Laurie Allen, Hannah Frost, Sarah Potvin, Elizabeth Russey Roke, and Stewart Varner. 2019. *Always Already Computational: Collections as Data.* 10.5281/zenodo.3152934. https://zenodo.org/records/3152935 Publisher: Zenodo.

[45] Thomas G. Padilla. 2018. Collections as data: Implications for enclosure. *College & Research Libraries News* 79, 6 (June 2018), 296. doi:10.5860/crln.79.6.296 Number: 6.

[46] Thomas G. Padilla and Devin Higgins. 2014. Library Collections as Humanities Data: The Facet Effect. *Public Services Quarterly* 10, 4 (Oct. 2014), 324–335. doi:10.1080/15228959.2014.963780 Publisher: Routledge _eprint: https://doi.org/10.1080/15228959.2014.963780.

[47] Katherine Panciera, Aaron Halfaker, and Loren Terveen. 2009. Wikipedians are born, not made: a study of power editors on Wikipedia. In *Proceedings of the 2009 ACM International Conference on Supporting Group Work (GROUP '09)*. Association for Computing Machinery, New York, NY, USA, 51–60. doi:10.1145/1531674.1531682

[48] Thomas Pellissier Tanon, Denny Vrandečić, Sebastian Schaffert, Thomas Steiner, and Lydia Pintscher. 2016. From Freebase to Wikidata: The Great Migration. In *Proceedings of the 25th International Conference on World Wide Web (WWW '16)*. International World Wide Web Conferences Steering Committee, Republic and Canton of Geneva, CHE, 1419–1428. doi:10.1145/2872427.2874809

[49] Kathleen Pine, Claus Bossen, Naja Holten Møller, Milagros Miceli, Alex Jiahong Lu, Yunan Chen, Leah Horgan, Zhaoyuan Su, Gina Neff, and Melissa Mazmanian. 2022. Investigating Data Work Across Domains: New Perspectives on the Work of Creating Data. In *CHI Conference on Human Factors in Computing Systems Extended Abstracts.* ACM, New Orleans LA USA, 1–6. doi:10.1145/3491101.3503724

[50] Alessandro Piscopo and Elena Simperl. 2018. Who Models the World? Collaborative Ontology Creation and User Roles in Wikidata. *Proc. ACM Hum.-Comput. Interact.* 2, CSCW (Nov. 2018), 141:1–141:18. doi:10.1145/3274410

[51] Jean-Christophe Plantin. 2019. Data Cleaners for Pristine Datasets: Visibility and Invisibility of Data Processors in Social Science. *Science, Technology, & Human Values* 44, 1 (Jan. 2019), 52–73. doi:10.1177/0162243918781268 Publisher: SAGE Publications Inc.

[52] Merrilee Proffitt. 2021. Leveraging Wikipedia: Connecting Communities of Knowledge. https://www.oclc.org/research/publications/2018/leveraging-wikipedia.html Journal Abbreviation: Leveraging Wikipedia: Connecting Communities of Knowledge Last Modified: 2022-8-3 Publisher: OCLC.

[53] Dirk Riehle, Philipp Riemer, Carsten Kolassa, and Michael Schmidt. 2014. Paid vs. Volunteer Work in Open Source. In *2014 47th Hawaii International Conference on System Sciences*. 3286–3295. doi:10.1109/HICSS.2014.407 ISSN: 1530-1605.

[54] Dana Rotman, Jenny Preece, Jen Hammock, Kezee Procita, Derek Hansen, Cynthia Parr, Darcy Lewis, and David Jacobs. 2012. Dynamic changes in motivation in collaborative citizen-science projects. In *Proceedings of the ACM 2012 conference on Computer Supported Cooperative Work (CSCW '12)*. Association for Computing Machinery, New York, NY, USA, 217–226. doi:10.1145/2145204.2145238

[55] Andrew L. Russell and Lee Vinsel. 2018. After Innovation, Turn to Maintenance. *Technology and Culture* 59, 1 (2018), 1–25. https://muse.jhu.edu/pub/1/article/692165 Publisher: Johns Hopkins University Press.

[56] Johnny Saldaña. 2013. *The coding manual for qualitative researchers* (2. ed ed.). SAGE Publ, Los Angeles, Calif.

[57] Amir Sarabadani, Aaron Halfaker, and Dario Taraborelli. 2017. Building Automated Vandalism Detection Tools for Wikidata. In *Proceedings of the 26th International Conference on World Wide Web Companion (WWW '17 Companion)*. International World Wide Web Conferences Steering Committee, Republic and Canton of Geneva, CHE, 1647–1654. doi:10.1145/3041021.3053366

[58] Michael J. Scroggins and Irene V. Pasquetto. 2020. Labor Out of Place: On the Varieties and Valences of (In)visible Labor in Data-Intensive Science. *Engaging Science, Technology, and Society* 6 (Jan. 2020), 111–132. doi:10.17351/ests2020.341

[59] Sina Semnani, Violet Yao, Heidi Zhang, and Monica Lam. 2023. WikiChat: Stopping the Hallucination of Large Language Model Chatbots by Few-Shot Grounding on Wikipedia. In *Findings of the Association for Computational Linguistics: EMNLP 2023*, Houda Bouamor, Juan Pino, and Kalika Bali (Eds.). Association for Computational Linguistics, Singapore, 2387–2413. doi:10.18653/v1/2023.findings-emnlp.157

[60] Priyanka Sen, Alham Fikri Aji, and Amir Saffari. 2022. Mintaka: A Complex, Natural, and Multilingual Dataset for End-to-End Question Answering. In *Proceedings of the 29th International Conference on Computational Linguistics*, Nicoletta Calzolari, Chu-Ren Huang, Hansaem Kim, James Pustejovsky, Leo Wanner, Key-Sun Choi, Pum-Mo Ryu, Hsin-Hsi Chen, Lucia Donatelli, Heng Ji, Sadao Kurohashi, Patrizia Paggio, Nianwen Xue, Seokhwan Kim, Younggyun Hahm, Zhong He, Tony Kyungil Lee, Enrico Santus, Francis Bond, and Seung-Hoon Na (Eds.). International Committee on Computational Linguistics, Gyeongju, Republic of Korea, 1604–1619. https://aclanthology.org/2022.coling-1.138

[61] Kartik Shenoy, Filip Ilievski, Daniel Garijo, Daniel Schwabe, and Pedro Szekely. 2022. A study of the quality of Wikidata. *Journal of Web Semantics* 72 (April 2022), 100679. doi:10.1016/j.websem.2021.100679

[62] Eira Tansey. 2016. Archives without archivists. *Reconstruction: Studies in Contemporary Culture* 16, 1 (March 2016). https://go.gale.com/ps/i.do?p=AONE&sw=w&issn=15474348&v=2.1&it=r&id=GALE%7CA484096660&sid=googleScholar&linkaccess=abs Publisher: Reconstruction.

[63] Divy Thakkar, Azra Ismail, Pratyush Kumar, Alex Hanna, Nithya Sambasivan, and Neha Kumar. 2022. When is Machine Learning Data Good?: Valuing in Public Health Datafication. In *CHI Conference on Human Factors in Computing Systems*. ACM, New Orleans LA USA, 1–16. doi:10.1145/3491102.3501868

[64] Bianca Trinkenreich, Mariam Guizani, Igor Wiese, Anita Sarma, and Igor Steinmacher. 2020. Hidden Figures: Roles and Pathways of Successful OSS Contributors. *Proc. ACM Hum.-Comput. Interact.* 4, CSCW2 (Oct. 2020), 180:1–180:22. doi:10.1145/3415251

[65] Janet Vertesi and Paul Dourish. 2011. The value of data: considering the context of production in data economies. In *Proceedings of the ACM 2011 conference on Computer supported cooperative work (CSCW '11)*. Association for Computing Machinery, New York, NY, USA, 533–542. doi:10.1145/1958824.1958906

[66] Veniamin Veselovsky, Dipto Sarkar, Jennings Anderson, and Robert Soden. 2022. An Automated Approach to Identifying Corporate Editing. *Proceedings of the International AAAI Conference on Web and Social Media* 16 (May 2022), 1052–1063. doi:10.1609/icwsm.v16i1.19357

[67] Nicholas Vincent, Isaac Johnson, and Brent Hecht. 2018. Examining Wikipedia With a Broader Lens: Quantifying the Value of Wikipedia's Relationships with Other Large-Scale Online Communities. In *Proceedings of the 2018 CHI Conference on Human Factors in Computing Systems*. ACM, Montreal QC Canada, 1–13. doi:10.1145/3173574.3174140

[68] Katy Weathington and Jed R. Brubaker. 2023. Queer Identities, Normative Databases: Challenges to Capturing Queerness On Wikidata. *Proceedings of the ACM on Human-Computer Interaction* 7, CSCW1 (April 2023), 1–26. doi:10.1145/3579517

[69] Diyi Yang, Aaron Halfaker, Robert Kraut, and Eduard Hovy. 2016. Who did what: Editor role identification in Wikipedia. In *Proceedings of the international AAAI conference on web and social media*, Vol. 10. 446–455. https://ojs.aaai.org/index.php/ICWSM/article/view/14732 Issue: 1.

[70] Yihan Yu and David W. McDonald. 2022. Unpacking Stitching between Wikipedia and Wikimedia Commons: Barriers to Cross-Platform Collaboration. *Proceedings of the ACM on Human-Computer Interaction* 6, CSCW2 (Nov. 2022), 1–35. doi:10.1145/3555766

[71] Charles Chuankai Zhang, Mo Houtti, C. Estelle Smith, Ruoyan Kong, and Loren Terveen. 2022. Working for the Invisible Machines or Pumping Information into an Empty Void? An Exploration of Wikidata Contributors' Motivations. *Proc. ACM Hum.-Comput. Interact.* 6, CSCW1 (April 2022), 135:1–135:21. doi:10.1145/3512982

[72] Charles Chuankai Zhang and Loren Terveen. 2021. Quantifying the Gap: A Case Study of Wikidata Gender Disparities. In *Proceedings of the 17th International Symposium on Open Collaboration (OpenSym '21)*. Association for Computing Machinery, New York, NY, USA, 1–12. doi:10.1145/3479986.3479992

[73] Xunhui Zhang, Tao Wang, Yue Yu, Qiubing Zeng, Zhixing Li, and Huaimin Wang. 2022. Who, What, Why and How? Towards the Monetary Incentive in Crowd Collaboration: A Case Study of Github's Sponsor Mechanism. In *Proceedings of the 2022 CHI Conference on Human Factors in Computing Systems (CHI '22)*. Association for Computing Machinery, New York, NY, USA, 1–18. doi:10.1145/3491102.3501822

[74] Yuxia Zhang, Mian Qin, Klaas-Jan Stol, Minghui Zhou, and Hui Liu. 2024. How Are Paid and Volunteer Open Source Developers Different? A Study of the Rust Project. In *Proceedings of the IEEE/ACM 46th International Conference on Software Engineering (ICSE '24)*. Association for Computing Machinery, New York, NY, USA, 1–13. doi:10.1145/3597503.3639197

[75] Haiyi Zhu, Robert E. Kraut, Yi-Chia Wang, and Aniket Kittur. 2011. Identifying shared leadership in Wikipedia. In *Proceedings of the SIGCHI Conference on Human Factors in Computing Systems (CHI '11)*. Association for Computing Machinery, New York, NY, USA, 3431–3434. doi:10.1145/1978942.1979453

[76] Lihong Zhu, Xu , Amanda, Deng , Sai, Heng , Greta, , and Xiaoli Li. 2023. Entity Management Using Wikidata for Cultural Heritage Information. *Cataloging & Classification Quarterly* 61, 1 (Jan. 2023), 20–46. doi:10.1080/01639374.2023.2188338 Publisher: Routledge _eprint: https://doi.org/10.1080/01639374.2023.2188338.

# 8 Appendix
## 8.1 Code Book

Table 2. CodeBook

| Code | Definition | What to Look For |
| --- | --- | --- |
| Data Curation Practices and Processes | Mentions of the what, the how, and the why work, include workflows and processes. Include comments about collaboration within the home institution and how they use Wikidata services to collaborate within their institutions. | Mentions of uploading dissertations and theses, creating namespaces for faculty members, and adding "archived at" locations. |
| Metadata Practices | Descriptions of specific metadata workflows, standards, or new data entities. Mentions of MARC. | Mentions of LOC name space authorities, confirming properties and objects, stories with acronyms, identifiers. |
| Community Norms | Mentions of Wikidata communities, (norms when contributing to wiki), multiple organizations (e.g; PCC), and home institution norms. | Descriptions how communities work together. |
| Institutional Integration & Tracking | Mentions of how Wikidata tools/resources are part of institutional workflows (leveraging wiki tools), and any information about how contributions are tracked. | Mentions of "Archived_at" tag as a way to promote institution access points, tracking how uploaded records are updated over time. |
| Training and Skills | Statements about the training and skills necessary to use Wikidata. | Examples of ad hoc training, skills set that are needed (SPARQL). |
| Authority and Expertise | Statements about being a 'good' editor/contributor, statements about proper use, best practices, or bad uses, and being credited for contributions on Wikidata. | For example, "it's not feasible to come up with best practices, every institution needs custom solutions". |
| Ethics, Values, and Passion projects | Statements about ethical uses, metadata justice, stories about passion projects and volunteerism. Stories about how their personal values shape their work on Wikidata. | Mentions of WikiProjects focusing on underrepresented populations; reactions to automated bot behavior. |
| Challenges and Barriers | Mentions of challenges to using Wikidata in memory institutions, and barriers to contributing to Wikidata. | For example, " Wikidata not always suitable for hosting indigenous knowledge". |
| Tooling | Mentions of SPARQL Service, OpenRefine, Quick Dtatements, Query Builder, word graph builders, and adding reference. Also include ILS/LMS names from institutions if mentioned. | Mentions of Voyager, Lux, SPRQL Service, Open Refine, other tools. |
| Whoa/Follow up | Interesting quotes or statements that aren't clearly understood. | |

## 8.2 Interview Protocol

**A. Background**

- May I confirm our records?
- What is your specific title/role? OR May I confirm that you are currently a ROLE/TITLE at memory Institution?
- How long have you been working at memory Institution?
- When did you graduate from MSIS/PHD/MLIS? From where? OR May I confirm that you graduated from UNIVERSITY with your ADVANCED DEGREE in Year?
- How is your role organized in your institution (for example, what team or unit are you in, and who do you report to)?

**Overview of Wikidata use in your organization**

- Can you describe how your organization uses Wikidata specifically?
- What teams are involved?
- How are tasks distributed across these teams?
- Why does it use Wikidata for these purposes?
- How did you start contributing to Wikidata? Is it part of your job requirement?
- The Wikimedia Foundation sees all its contributors as "volunteers". Do you see your work on Wikidata as volunteer work? Do you identify yourself as a volunteer for Wikidata?

**Contributing to Wikidata**

- This next section of questions is about contributing data to Wikidata. Do you have specific contribution experiences?
- Can you walk me through a typical task you would do on Wikidata?
- Do you mind sharing your username with us so we can retrieve your contribution history for further analysis?
- Are there any tools you use in contributing to Wikidata? And why?
- Could you walk us through an example?
- How do you decide what kind of records to upload to Wikidata?
- Does your organization have a particular focus on what type of contributions to make on Wikidata?
- What obstacles do you face when contributing to Wikidata?
- What specific GLAM tools would you like to have to contribute more to Wikidata?

**D. Monitoring and Tracking**

- Do you care about being credited when you contribute to Wikidata? For example, would you like to be a named Creator of Wikidata records? Or would you prefer for it to be credited to your institution?
- Do you or your organization track your edits on achieving the organizational goal?
- Who do you envision benefiting from your edits and how?

**E. External Collaboration**

- How do you collaborate with other volunteers on Wikidata?
- Have you encountered instances where there are conflicting contributions? This could be with bots or with real Wikidata volunteers.
- How about conflicting views? This could be when you created an entry that does not meet Wikidata's notability requirements or when your source was questioned by others.
- How do you collaborate with other memory institutions on Wikidata?

**F. Querying Wikidata**

- What existing query services do you or your organization use? (e.g., do you use SPARQL)
- What kind of queries do you or your organization typically craft or use?
- What are the limitations of existing Wikidata query services that you encounter?

**G. Misc**
- Are there other questions that I should ask or that you want to discuss?